\documentstyle[twocolumn,aps,epsfig]{revtex}
\begin{document}
\draft
\twocolumn[\hsize\textwidth\columnwidth\hsize\csname @twocolumnfalse\endcsname
\title{ Magnetic phase diagram of doped CMR manganites }
\author{Unjong Yu, Yookyung Jo, B. I. Min}
\address{Department of Physics,
        Pohang University of Science and Technology, 
        Pohang 790-784, Korea}
\date{\today}
\maketitle
\begin{abstract}
The magnetic phase diagram of the colossal magnetoresistance (CMR)
manganites is determined 
based on the Hamiltonian incorporating the double-exchange (DE) 
interaction between degenerate Mn $e_g$ orbitals and the antiferromagnetic
(AF) superexchange interaction between Mn $t_{2g}$ spins.
We have employed the rigorous quantum mechanical formalism and obtained 
the finite temperature phase diagram which describes well the
commonly observed features in CMR manganites.
We have also shown that the CE-type AF 
structure cannot be stabilized at $x$=0.5 in this model.
\end{abstract}
\pacs{{\bf Keywords:} CMR-manganites; Phase diagram; 
                     Double-exchange; Superexchange }
\vskip2pc]
\narrowtext
Colossal magneto-resistance (CMR) manganites R$_{1-x}$A$_x$MnO$_3$ 
(R = rare-earth; A = divalent cation) have attracted much
attention due to many unusual 
physical properties and complicated phase diagrams.
\cite{Schiffer}. 
The generic magnetic structure in the ground state is found to be 
A$\rightarrow$F$\rightarrow$A$\rightarrow$C$\rightarrow$G-type with 
hole doping.
A, C, and G-type denote layered type, rod type, and NaCl type antiferromagnetic
(AF) structures, respectively.
The major mechanism determining the magnetic structure of these compounds 
are the double-exchange (DE) which induces metallic ferromagnetism \cite{Zener}
and AF superexchange (SE) interactions. 

The combined Hamiltonian of the DE and SE interactions is given by
\begin{eqnarray}
H = - \!\sum_{\langle i,j\rangle \alpha \beta} t_0^{\alpha \beta} \, 
           \cos\left(\frac{\theta_{ij}}{2}\right) \,
       c_{i \alpha}^{\dagger} c_{j \beta}^{}
    + J \sum_{\langle i,j\rangle} {\bf S}_{i} \cdot {\bf S}_{j}.
\end{eqnarray}
Here, $t_0$ and $J$ are hopping and AF parameters, respectively.
$\langle i,j\rangle$ denotes the nearest neighbor pairs and ${\bf S}_{i}$ 
is the local $t_{2g}$ spin of the $i$th site. $\alpha$ and $\beta$ 
represent the two orbitals of Mn $e_g$ band. $\theta_{ij}$
amounts to the distortion angle of spins at the 
sites $i$ and $j$. Brink and Khomskii\cite{Brink} solved this Hamiltonian 
with the anisotropic mean field approximation in the classical limit. 
We have refined this approach; (i) by using the  quantum
mechanical treatment
instead of taking the classical limit,
and (ii) by generalizing the solution at the finite temperature.

The quantum mechanical expression of $\cos(\theta_{ij}/2)$ is 
$(\hat{S}_0+1/2)/(2S+1)$, where $S$ is the magnitude of local $t_{2g}$
spin ($S=3/2$), and $\hat{S}_0$ is the total spin of the sites $i$ and
$j$ \cite{Anderson}.
We adopted three order parameters, $\lambda$, $\Theta_{xy}$, and 
$\Theta_{z}$. $\lambda$ is the mean field determining whether 
the system is magnetically ordered or not, and $\Theta_{xy}$ and 
$\Theta_z$ are the average relative spin angles between neighboring sites 
in the x-y plane and along the z-direction, respectively.

With these order parameters, the free energy per site can be comprised 
as follows.
\begin{eqnarray}
\frac{1}{N} F(\lambda, && \Theta_{xy}, \Theta_{z}, T, y) = \nonumber \\
  &&E_{\rm band} (\lambda, \Theta_{xy}, \Theta_{z}, y)
    + E_{\rm AF} (\lambda, \Theta_{xy}, \Theta_{z}) \nonumber \\
  &&- \ T \left(\, y \, \sigma(\lambda, S=2) 
    + (1-y) \, \sigma(\lambda, S=3/2)\, \right).
\end{eqnarray}
$E_{\rm band}$ can be obtained by filling up the states with the electron 
number density $y$ ($y = 1-x$).
$E_{\rm AF}$ denotes the SE energy 
and the entropy is given by $\sigma(\lambda, S) = \log \nu - 
\lambda \langle S_z\rangle$, where $\nu$ is the partition function, and 
$\langle S_z \rangle$ is an expectation value of spin component 
in the direction of the mean field $\bbox{\lambda}$ \cite{Kubo}. 
By minimizing this free 
energy with respect to three parameters,
one can determine the equilibrium phases.
We have performed the rigorous quantum 
mechanical calculations for $(\hat{S}_0+1/2)/(2S+1)$ and 
$\langle{\bf S}_i \cdot {\bf S}_j\rangle$.

\begin{figure}[!t]
\centerline{\epsfig{figure=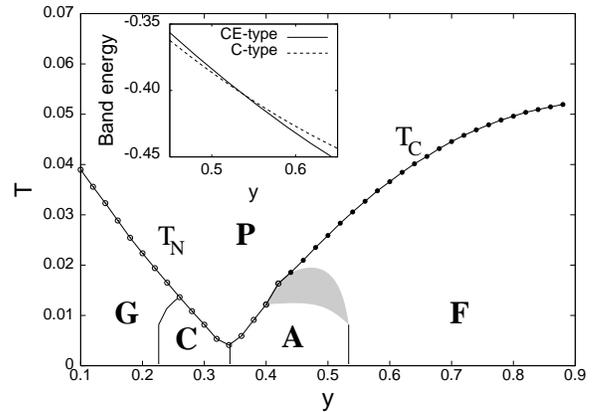,width=8.0cm}}
\caption{Finite temperature phase diagram of doped manganites
for $J/t_0=0.018$. The gray region represents a region where the stability
between A and F states is indeterminable.
(inset) Band energies of the CE and C types with respect to electron doping.}
\label{phase}
\end{figure}

  The finite temperature phase diagram is given in Fig.~\ref{phase} for 
$J/t_0 = 0.018$. At high temperature, the system becomes paramagnetic.
At low temperature, the G$\rightarrow$C$\rightarrow$A$\rightarrow$F-type 
magnetic transitions appear with electron doping. $T_{\rm N}$($T_{\rm C}$) 
first 
decreases with electron doping and hits its minimum at the C$\rightarrow$A
phase transition point and then increases. This magnetic structure change
and $T_{\rm N}$ minimum at C$\rightarrow$A phase transition point coincide 
with the experimental observations in Nd$_{1-x}$Sr$_{x}$MnO$_3$ \cite{Tokura}.
Also notable is that the A-type AF ground state experiences a 
ferromagnetic transition in the doping region close to the ferromagnetic 
ground state before it becomes a paramagnetic state at high temperature.
Indeed, this phenomenon is observed\cite{Akimoto}.

We have calculated the ground state energy of the CE-type\cite{Brink2} 
AF structure
as a function of doping. 
Because the C and CE-type have the 
same magnetic energy, the band energy alone determines the preferable 
ground state. 
From the inset of Fig.~\ref{phase},
one can verify that the CE-type has the lower band 
energy than the C-type for $y > 0.54$ ($x < 0.46$).
Thus, it is impossible to explain the CE-type ground state near $x=0.5$  
within this model. Prior to this result, Brink {\it et al}.\cite{Brink2} 
performed the same calculation in the classical limit. According to their 
result, the CE-type has the lower energy than the C-type for $x<0.57$ and 
they concluded that the CE-type naturally emerges near $x=0.5$ in the
same model.
Our results, however, indicates that their classical approximation 
fails at this point and that other mechanism such as charge, orbital ordering 
and structural distortion are required to explain the CE-type ground state.
Finally, it should be pointed out that the description of the A-type AF 
near $x=0$ is beyond the scope of this model. 
In that case, the orbital degeneracy
and the Jahn-Teller effect are thought to be crucial.

In conclusion, with the quantum mechanical calculation of the combined
model of the DE and SE interactions, we have obtained the finite temperature
phase diagram of doped perovskite manganites.
We have also shown that the CE-type AF structure 
usually observed near $x=0.5$ cannot be explained within this
model. 

Acknowledgments$-$
This work was supported by the KRF grant (KRF-2002-070-C00038).

\newpage

\end{document}